\documentclass{pasj00}
\begin{document}
\SetRunningHead{J.\ Fukue}
{Relativistic Radiative Flow in a Luminous Disk}
\Received{2005/06/24}
\Accepted{2005/10/06}

\title{Relativistic Radiative Flow in a Luminous Disk}

\author{Jun \textsc{Fukue}} 
\affil{Astronomical Institute, Osaka Kyoiku University, 
Asahigaoka, Kashiwara, Osaka 582-8582}
\email{fukue@cc.osaka-kyoiku.ac.jp}



\maketitle


\begin{abstract}
Radiatively driven transfer flow perpendicular to a luminous disk
was examined under a fully special relativistic treatment,
taking into account radiation transfer.
The flow was assumed to be vertical, and
the gravity, the gas pressure, and the viscous heating were ignored.
In order to construct the boundary condition at the flow top,
the magic speed above the flat source was re-examined,
and it was found that
the magic speed above a moving source can exceed
that above a static source ($\sim 0.45~c$).
Then, the radiatively driven flow in a luminous disk
was numerically solved, 
from the flow base (disk ``inside''), where the flow speed is zero, 
to the flow top (disk ``surface''), where the optical depth is zero.
For a given optical depth and appropriate initial conditions
at the flow base, where the flow starts,
a loaded mass in the flow was obtained as an eigenvalue
of the boundary condition at the flow top.
Furthermore, a loaded mass and the flow final speed at the flow top
were obtained as a function of the radiation pressure at the flow base;
the flow final speed increases as the loaded mass decreases.
Moreover, the flow velocity and radiation fields along the flow
were obtained as a function of the optical depth.
Within the present treatment,
the flow three velocity $v$ is restricted to be within the range of
$v < c/\sqrt{3}$, which is the relativistic sound speed,
due to the relativistic effect.
\end{abstract}

\section{Introduction}

Accretion disks are tremendous energy sources in the active universe
(see Kato et al. 1998 for a review).
In particular, when the mass-accretion rate highly exceeds the critical rate, 
the disk local luminosity exceeds the Eddington one,
and mass loss from the disk surface 
driven by radiation pressure takes place.

Such a radiatively driven outflow from a luminous disk
has been extensively studied in the context of
models for astrophysical jets by many researchers
(Bisnovatyi-Kogan, Blinnikov 1977; Katz 1980; Icke 1980; Melia, K\"onigl 1989; 
Misra, Melia 1993; Tajima, Fukue 1996, 1998; Watarai, Fukue 1999;
Hirai, Fukue 2001; Fukue et al. 2001; Orihara, Fukue 2003),
as on-axis jets (Icke 1989; Sikora et al. 1996; Renaud, Henri 1998;
Luo, Protheroe 1999; Fukue 2005a),
as outflows confined by a gaseous torus
(Lynden-Bell 1978; Davidson, McCray 1980; Sikora, Wilson 1981; Fukue 1982),
or as jets confined by the outer flow or corona
(Sol et al. 1989; Marcowith et al. 1995; Fukue 1999),
and as numerical simulations
(Eggum et al. 1985, 1988).
In almost all of these studies, however,
the disk radiation fields were treated as external fields,
and the radiation transfer was not solved.

The radiation transfer in the disk, on the other hand,
was investigated in relation to the structure
of a static disk atmosphere and
the spectral energy distribution from the disk surface
(e.g., Meyer, Meyer-Hofmeister 1982; Cannizzo, Wheeler 1984;
Shaviv, Wehrse 1986; Adam et al. 1988;
Hubeny 1990; Mineshige, Wood 1990; Ross et al. 1992; Artemova et al. 1996;
Hubeny, Hubeny 1997, 1998; Hubeny et al. 2000, 2001;
Davis et al. 2005; Hui et al. 2005).
In these studies, however,
the vertical movement and mass loss were not considered.
Moreover, their treatments were restricted
in the non-relativistic regime, and
the relativistic effects were not considered.

In order to break such a situation, where
radiation transfer has not been considered
in the radiatively driven wind from the disk,
we recently examined the radiatively driven vertical outflow
-- {\it moving photosphere} -- 
in a luminous flat disk (Fukue 2005b, c).
In these papers,
the radiative transfer flow was analytically or numerically solved,
but the obtained flow speed was limited in the subrelativistic regime,
since the problem was treated up to the order of $(v/c)^1$.
In some astrophysical jet sources, however,
the jet speed is mildly or highly relativistic.
In gamma-ray bursts,
the flow speed is supposed to be extremely relativistic.
Hence, in the next step, we should consider
the transfer flow up to the full order of $(v/c)^2$
(cf. Fukue 1999; Hirai, Fukue 2001;
Fukue et al. 2001; Orihara, Fukue 2003; Fukue 2005a
without transfer).
Moreover, although it was incorporated 
in the previous paper (Fukue 2005b), which were
up to the first order of the flow velocity,
the effect of radiation drag must become much more important
(Phinney 1987; Icke 1989).
Other relativistic effects would further appear
in the fully relativistic regime.

In this paper,
we thus examine the radiatively driven transfer outflow
in a luminous flat disk
within the fully relativistic regime of $(v/c)^2$.
At the preliminary stage, 
we ignore the gravity of the central object, the gas pressure, 
and the viscous heating,
and we treat the wind as a vertical one-dimensional flow
without rotational motion.

In the next section
we describe basic equations in the vertical direction.
In section 3
we examine the boundary condition at the flow top,
and show the magic speed above the moving source.
In section 4
we then solve and examine a radiative flow
under the appropriate boundary conditions
at the flow base and top.
The final section is devoted to concluding remarks.


\section{Basic Equations}

Let us suppose a luminous flat disk,
deep inside which
gravitational or nuclear energy is released
via viscous heating or other processes.
The radiation energy is transported in the vertical direction,
and the disk gas, itself, also {\it moves} in the vertical direction
due to the action of radiation pressure (i.e., plane-parallel approximation).
For the sake of simplicity, in the present paper,
the radiation field is considered to be sufficiently intense that
both the gravitational field of, e.g., the central object
and the gas pressure can be ignored:
tenuous cold normal plasmas in the super-Eddington disk,
cold pair plasmas in the sub-Eddington disk, or
dusty plasmas in the sub-Eddington disk.
Internal heating is also ignored:
the flow in or near to the surface envelope of the disk.
As for the order of the flow velocity $v$,
we consider the fully special relativistic regime,
where the terms are retained up to the second order of $(v/c)$.

Under these assumptions,
the radiation hydrodynamic equations
for steady vertical ($z$) flows are described as follows
(Kato et al. 1998).

The continuity equation is
\begin{equation}
   \rho cu = J ~(={\rm const.}),
\label{rho1}
\end{equation}
where $\rho$ is the proper gas density, $u$ the vertical four velocity, 
$J$ the mass-loss rate per unit area,
and $c$ the speed of light.
The four velocity $u$ is related to the three velocity $v$ by
$u=\gamma v/c$, where $\gamma$ is the Lorentz factor:
$\gamma=\sqrt{1+u^2}=1/\sqrt{1-(v/c)^2}$.

The equation of motion is
\begin{equation}
   c^2u\frac{du}{dz} = \frac{\kappa_{\rm abs}+\kappa_{\rm sca}}{c}
                    \left[ F \gamma (1+2u^2) - c(E+P)\gamma^2 u \right],
\label{u1}
\end{equation}
where $\kappa_{\rm abs}$ and $\kappa_{\rm sca}$
are the absorption and scattering opacities (gray),
defined in the comoving frame,
$E$ the radiation energy density, $F$ the radiative flux, and
$P$ the radiation pressure observed in the inertial frame.
The first term in the brackets on the right-hand side
of equation (\ref{u1}) means the radiatively driven force,
which is modified to the order of $u^2$, whereas
the second term is the radiation drag force,
which is also modified, but roughly proportional to the velocity.

In the no-gas pressure approximation and without heating,
the energy equation is reduced to a radiative equilibrium relation,
\begin{equation}
   0 =  j - c\kappa_{\rm abs} E \gamma^2 - c\kappa_{\rm abs} P u^2
                  + 2 \kappa_{\rm abs} F \gamma u,
\label{j1}
\end{equation}
where $j$ is the emissivity defined in the comoving frame.
In this equation (\ref{j1}),
the third and fourth terms on the right-hand side
appear in the relativistic regime.

For radiation fields, the zeroth-moment equation becomes
\begin{eqnarray}
   \frac{dF}{dz} &=& \rho \gamma
         \left[ j - c\kappa_{\rm abs} E
                 + c\kappa_{\rm sca}(E+P)u^2  \right.
\nonumber
\\
    &&   \left. + \kappa_{\rm abs}Fu/\gamma
               -\kappa_{\rm sca}F ( 1+v^2/c^2 )\gamma u \right].
\label{F1}
\end{eqnarray}
The first-moment equation is
\begin{eqnarray}
   \frac{dP}{dz} &=& \frac{\rho \gamma}{c} 
         \left[ ju/\gamma - \kappa_{\rm abs} F
                  + c\kappa_{\rm abs}Pu/\gamma \right.
\nonumber
\\
     && \left. -\kappa_{\rm sca}F(1+2u^2)
               +c\kappa_{\rm sca}(E+P)\gamma u \right].
\label{P1}
\end{eqnarray}

Finally, the closure relation {\it in the inertial frame} is
\begin{equation}
   cP \left( 1 + \frac{2}{3}u^2 \right) = 
   cE \left( \frac{1}{3} - \frac{2}{3} u^2 \right) 
   + \frac{4}{3} F \gamma u.
\label{E}
\end{equation}
As a closure relation, 
the usual Eddington approximation {\it in the comoving frame} is adopted.
Radiative quantities are then transformed from the comoving frame
to the inertial frame, and we have the closure relation (\ref{E})
in the inertial frame (see Kato et al. 1998 for details).

Eliminating $j$ with the help of equations (\ref{j1}),
and using continuity equation (\ref{rho1}),
equations (\ref{u1}), (\ref{F1}), and (\ref{P1}) are rearranged as
\begin{eqnarray}
   cJ\frac{du}{dz} &=& (\kappa_{\rm abs}+\kappa_{\rm sca})
                     \rho \frac{\gamma}{c}
                    \left[ F (1+2u^2) - c(E+P)\gamma u \right],
\label{u2}
\\
   \frac{dF}{dz} &=& (\kappa_{\rm abs}+\kappa_{\rm sca})
                   \rho u
                    \left[ c(E+P)\gamma u - F (1+2u^2) \right],
\label{F2}
\\
   \frac{dP}{dz} &=& (\kappa_{\rm abs}+\kappa_{\rm sca})
                    \rho \frac{\gamma}{c} 
                    \left[ c(E+P)\gamma u - F (1+2u^2) \right].
\label{P2}
\end{eqnarray}

Integrating the sum of equations (\ref{u2}) and (\ref{P2})
yields to the momentum flux conservation along the flow,
\begin{equation}
   cJ u + P = K ~(={\rm const.}).
\label{K}
\end{equation}
In the subrelativistic regime,
this relation is reduced to that derived in Fukue (2005b).
Similarly, after some manipulations,
integrating the sum of equations (\ref{u2}) and (\ref{F2})
gives the energy flux conservation along the flow,
\begin{equation}
   c^2 J \gamma + F = L ~(={\rm const.}).
\label{L}
\end{equation}
In the subrelativistic regime,
this relation means that the flux $F$ is constant.

At this stage, the basic equations are
the equation of motion (\ref{u2}),
the mass flux (\ref{rho1}), the momentum flux (\ref{K}),
the energy flux (\ref{L}), and the closure relation (\ref{E}).

Next, by introducing the optical depth $\tau$ by
\begin{equation}
    d\tau = - ( \kappa_{\rm abs}+\kappa_{\rm sca} ) \rho dz,
\end{equation}
the equation of motion (\ref{u2}) is rewritten as
\begin{equation}
   cJ\frac{du}{d\tau} = - \frac{\gamma}{c}
                    \left[ F (1+2u^2) - c(E+P)\gamma u \right].
\label{u3}
\end{equation}
Furthermore, eliminating $E$ with the help of equation (\ref{E}),
this equation (\ref{u3}) can be finally rewritten as
\begin{equation}
   cJ\frac{du}{d\tau} = -\frac{\gamma}{c}
           \frac{ F(1+4u^2) - 4cP \gamma u}{1-2u^2},
\label{u}
\end{equation}
or
\begin{equation}
   c^2 J \gamma^2 \frac{d\beta}{d\tau} = 
           - \frac{ F(1+3\beta^2) - 4cP \beta}{1-3\beta^2},
\label{beta}
\end{equation}
where $\beta=v/c$.

We shall solve equations (\ref{u}), (\ref{K}), and (\ref{L})
for appropriate boundary conditions.
Before this,
we discuss the boundary conditions at the flow top
in the next section.

\section{Magic Speed Above a Moving Photosphere}

When there is no motion in a luminous flat disk (``static photosphere''),
the radiation fields above the disk are easily obtained.
Namely, just above the disk with surface intensity $I_0$,
the radiation energy density $E_{\rm s}$, 
the radiative flux $F_{\rm s}$, and
the radiation pressure $P_{\rm s}$ are
$(2/c)\pi I_0$,
$\pi I_0$, and $(2/3c)\pi I_0$, respectively,
where subscript s denotes the values at the disk surface.
In the problem of radiation transfer in the accretion disk,
these values are usually adopted as boundary conditions
(e.g., Artemova et al. 1996).
As will be shown below, however,
the radiation fields above the luminous disk are changed
when the disk gas itself does move upward (``moving photosphere'').
Even in such a case, however,
if the flow speed is small compared with the speed of light,
the conditions for a static photosphere would be approximately adopted,
and we can use these conditions in a previous paper (Fukue 2005b),
where the flow speed is limited in the subrelativistic regime.
When the flow speed is of the order of the speed of light,
on the other hand,
we should carefully treat the boundary condition
for the moving photosphere.
Thus, in the present paper, where the flow is treated
in a fully relativistic manner,
we must derive the exact boundary conditions
above the moving photosphere.

In addition, for the radiatively driven flow in the relativistic regime,
it becomes important the effect of radiation drag,
which suppresses the jet speed (Phinney 1987; Icke 1989).
For the flow above the static photosphere without gravity, 
Icke (1989) found that the {\it magic speed} of jets 
becomes $[(4-\sqrt{7})/3]~c \sim 0.45~c$.
If the photophere is moving, however,
such a magic speed will be also changed (cf. Fukue 2000).

Hence, before we can examine the relativistic radiative flow,
we must derive the radiation fields above the moving photosphere
and consider the boundary condition at the flow top (disk ``surface'').

Let us suppose the situation that
a flat infinite photosphere is not static,
but moving upward with a speed $v_{\rm s}$ 
($=c\beta_{\rm s}$, and
the corresponding Lorentz factor is $\gamma_{\rm s}$).
Then, the direction and intensity of radiation
are changed due to aberration and Doppler effects 
(cf. Kato et al. 1998; Fukue 2000).

The transformations of the photon frequency $\nu$
and photon direction $\theta$
between the inertial and comoving frames become
\begin{eqnarray}
   \frac{\nu_0}{\nu} &=& 
   {\gamma_{\rm s} \left( 1- \beta_{\rm s} \cos\theta \right)}
   = \frac{1}{\gamma_{\rm s} \left( 1+ \beta_{\rm s} \cos\theta_0 \right)},
\label{doppler}
\\
   \cos\theta &=&
   \frac{\cos\theta_0 + \beta_{\rm s}}{1 + \beta_{\rm s} \cos\theta_0},
\label{aberration}
\end{eqnarray}
where subscript 0 means the values
measured in the comoving frame and
$\gamma_{\rm s} = 1/\sqrt{1-\beta_{\rm s}^2}$.
For incident radiation
with $\theta_0=\pi/2$ in the comoving frame,
the direction cosine in the inertial frame is
$\cos\theta=\beta_{\rm s}$.

Furthermore, 
the transformation of the intensity $I$
between the inertial and comoving frames is
\begin{eqnarray}
   I_0 = \left( \frac{\nu_0}{\nu} \right)^4 I
       &=& 
  {\left[\gamma_{\rm s} \left( 1- \beta_{\rm s} \cos\theta \right)\right]^4}I
\nonumber \\
       &=&
  \frac{1}
  {\left[\gamma_{\rm s} \left( 1+\beta_{\rm s} \cos\theta_0 \right)\right]^4}I.
\label{intensity}
\end{eqnarray}

\begin{figure}
  \begin{center}
  \FigureFile(80mm,80mm){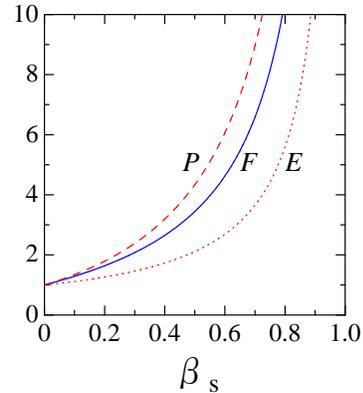}
  \end{center}
\caption{
Radiation fields above a moving photosphere with speed $v_{\rm s}$
($=c\beta_{\rm s}$).
Each component of the radiation fields is normalized
by the values of a static case.
Every component increases as the speed increases.
}
\end{figure}

Considering the Doppler effect (\ref{doppler}) 
and aberration (\ref{aberration}),
the radiation fields above the moving photosphere 
in the inertial frame are calculated as follows:
\begin{eqnarray}
   cE_{\rm s} &=& \int_0^{\cos^{-1}\beta_{\rm s}} I d\Omega
\nonumber \\
   &=& \frac{2\pi I_0}{\gamma_{\rm s}^4}
       \int_0^{\cos^{-1}\beta_{\rm s}} 
              \frac{\sin\theta d\theta}
                   {\left( 1-\beta_{\rm s} \cos\theta \right)^4}
\nonumber \\
   &=& {2\pi I_0}{\gamma_{\rm s}^2}
       \frac{3+3\beta_{\rm s}+\beta_{\rm s}^2}{3}
\nonumber \\
   &=& {2\pi I_0}
       \frac{3\gamma_{\rm s}^2+3\gamma_{\rm s}u_{\rm s}+u_{\rm s}^2}{3},
\label{Es}
\end{eqnarray}
\begin{eqnarray}
   F_{\rm s} &=& \int_0^{\cos^{-1}\beta_{\rm s}} I \cos\theta d\Omega
\nonumber \\
   &=& \frac{2\pi I_0}{\gamma_{\rm s}^4}
       \int_0^{\cos^{-1}\beta_{\rm s}} 
              \frac{\sin\theta \cos\theta d\theta}
                   {\left( 1-\beta_{\rm s} \cos\theta \right)^4}
\nonumber \\
   &=& {2\pi I_0}{\gamma_{\rm s}^2}
       \frac{3+8\beta_{\rm s}+3\beta_{\rm s}^2}{6}
\nonumber \\
   &=& {2\pi I_0}
       \frac{3\gamma_{\rm s}^2+8\gamma_{\rm s}u_{\rm s}+3u_{\rm s}^2}{6},
\label{Fs}
\end{eqnarray}
\begin{eqnarray}
   cP_{\rm s} &=& \int_0^{\cos^{-1}\beta_{\rm s}} I \cos^2\theta d\Omega
\nonumber \\
   &=& \frac{2\pi I_0}{\gamma_{\rm s}^4}
       \int_0^{\cos^{-1}\beta_{\rm s}} 
              \frac{\sin\theta \cos^2\theta d\theta}
                   {\left( 1-\beta_{\rm s} \cos\theta \right)^4}
\nonumber \\
   &=& {2\pi I_0}{\gamma_{\rm s}^2}
       \frac{1+3\beta_{\rm s}+3\beta_{\rm s}^2}{3}
\nonumber \\
   &=& {2\pi I_0}
       \frac{\gamma_{\rm s}^2+3\gamma_{\rm s}u_{\rm s}+3u_{\rm s}^2}{3}.
\label{Ps}
\end{eqnarray}
That is,
the radiation fields depend on the speed $v_{\rm s}$ of the photosphere,
and every component increases as the speed increases (see figure 1).

We must use these values of radiation fields
as boundary conditions above the moving photosphere.

\begin{figure}
  \begin{center}
  \FigureFile(80mm,80mm){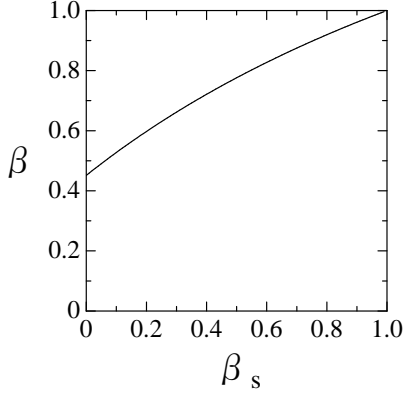}
  \end{center}
\caption{
Magic speed $\beta$ above a moving photosphere
as a function of the speed $\beta_{\rm s}$
of the photosphere (disk ``surface'').
When the photosphere is static,
the magic speed is $\sim 0.45c$.
When the photosphere is moving, however,
the magic speed can exceed this limit.
}
\end{figure}

In addition, 
the magic speed above the luminous infinite disk 
is obtained by the condition
where the radiative force is balanced with the radiation drag force.
Hence, from equation (\ref{u1}), we have
\begin{eqnarray}
   0 &=&  F_{\rm s} (1+2u^2) - c(E_{\rm s}+P_{\rm s})\gamma u
\nonumber \\
     &=&  [F_{\rm s} (1+\beta^2) - c(E_{\rm s}+P_{\rm s})\beta]\gamma^2,
\label{magic}
\end{eqnarray}
for the magic speed $\beta$ ($=u/\gamma$).

Inserting equations (\ref{Es})--(\ref{Ps}) into this equation (\ref{magic}),
we can obtain the {\it magic speed above a moving photosphere}
as a function of the speed $\beta_{\rm s}$ of the photosphere,
\begin{equation}
   \beta=\frac{4-\sqrt{7}+6\beta_{\rm s}+(4+\sqrt{7})\beta_{\rm s}^2}
              {3+8\beta_{\rm s}+3\beta_{\rm s}^2}.
\end{equation}
When the disk is static ($\beta_{\rm s}=0$),
this relation is reduced to that obtained by Icke (1989),
$\beta=(4-\sqrt{7})/3 \sim 0.45$.

In figure 2,
the magic speed above a moving photosphere is shown
as a function of the speed of the photosphere.
When the luminous photosphere is not static, but moving,
the magic speed can exceed the limit of $\sim 0.45c$,
which Icke (1989) obtained for a static photosphere (cf. Fukue 2000).

\section{Relativistic Radiative Transfer Flow}

Now, we discuss our numerical solution
of equations (\ref{u}), (\ref{K}), and (\ref{L})
for appropriate boundary conditions.

\subsection{Boundary Conditions and Singularity}

As for boundary conditions,
we impose the following cases.
At the flow base (disk ``inside'')
with an arbitrary optical depth $\tau_0$
(which relates to the disk surface density),
the flow velocity $u$ is zero,
the radiative flux is $F_0$
(which is a measure of the strength of radiation field), and
the radiation pressure is $P_0$
(which connects with the radiation pressure gradient
and relates to the disk internal structure),
where subscript 0 denotes the values at the flow base.
At the flow top (disk ``surface'')
where the optical depth is zero,
the radiation fields should satisfy the values
above a moving photosphere derived in the previous section.
Namely, just above the disk with surface intensity $I_{\rm s}$,
the radiation energy density $E_{\rm s}$, 
the radiative flux $F_{\rm s}$, and
the radiation pressure $P_{\rm s}$ are, respectively,
\begin{eqnarray}
   cE_{\rm s} 
   &=& {2\pi I_{\rm s}}
       \frac{3\gamma_{\rm s}^2+3\gamma_{\rm s}u_{\rm s}+u_{\rm s}^2}{3},
\label{Es2}
\\
   F_{\rm s}
   &=& {2\pi I_{\rm s}}
       \frac{3\gamma_{\rm s}^2+8\gamma_{\rm s}u_{\rm s}+3u_{\rm s}^2}{6},
\label{Fs2}
\\
   cP_{\rm s}
   &=& {2\pi I_{\rm s}}
       \frac{\gamma_{\rm s}^2+3\gamma_{\rm s}u_{\rm s}+3u_{\rm s}^2}{3},
\label{Ps2}
\end{eqnarray}
where $u_{\rm s}$ ($=\gamma_{\rm s}v_{\rm s}/c$)
is the flow four velocity at the flow top and
subscript s denotes the values at the flow top.

Applying these boundary conditions to equations (\ref{K}) and (\ref{L}),
we have two relations on the boundary values and mass-loss rate:
\begin{eqnarray}
   Jc^2 u_{\rm s} + cP_{\rm s} &=& cP_0,
\label{bc1}
\\
   Jc^2 \gamma_{\rm s} + F_{\rm s} &=& Jc^2 + F_0.
\label{bc2}
\end{eqnarray}

Physically speaking,
in the radiative flow starting from the flow base
with an arbitrary optical depth $\tau_0$,
for initial values of $F_0$ and $P_0$ at the flow base,
the final values of
the radiation fields $E_{\rm s}$, $F_{\rm s}$, $P_{\rm s}$, and
the flow velocity $u_{\rm s}$ at the flow top
can be obtained by solving basic equations.
Furthermore, the mass-loss rate $J$
is determined as an eigenvalue
so as to satisfy the bondary condition at the flow top
(cf. Fukue 2005b in the subrelativistic regime).

In the present full relativistic case, however,
the final values of the radiation fields at the flow top depend
on the flow velocity there,
and the final values at the flow top
cannot be analytically expressed by the initial values at the flow base.
Hence, in this paper
we determine the mass-loss rate as follows.

In radiative flow with optical depth $\tau_0$,
we first give the final flow velocity $u_{\rm s}$ (and $\gamma_{\rm s}$),
instead of the initial value of $P_0$.
Then, the final values of radiation fields $E_{\rm s}$, $F_{\rm s}$,
and $P_{\rm s}$ can be fixed by equations (\ref{Es2})--(\ref{Ps2}).
Next, we give a trial value for the mass-loss rate $J$,
and the initial values of $P_0$ and $F_0$
can be fixed by equations (\ref{bc1}) and (\ref{bc2}).
Since all the parameters are temporarily fixed,
we solve equation (\ref{u}) from $\tau=\tau_0$ to $\tau=0$.
Generally, however, the obtained final velocity at $\tau=0$
is different from a given $u_{\rm s}$.
Thus, we vary the value of $J$
and follow iterative processes,
so that the calculated final velocity coincides 
with a given final velocity $u_{\rm s}$.

Another point to be noticed is the {\it singularity}
in equation (\ref{u}):
i.e.,
the denominator of equation (\ref{u}) vanishes
when $u=\pm 1/\sqrt{2}$ or $\beta=\pm 1/\sqrt{3}$.
This singularity originates from the closure relation (\ref{E}),
and coincides with the {\it sound speed}
of the relativistic (photon) gas.
In other words,
equation (\ref{u}) has a form
of the transonic wind equation for the relativistic (photon) gas
with sound speed $c/\sqrt{3}$.
In usual wind equations from a gravitating source,
there exist transonic (critical) points,
where both the numerator and denominator of wind equations
vanish simultaneously.
In the present case,
there also exist {\it critical points},
which yield that
$u_{\rm c}=\pm 1/\sqrt{2}$ ($\beta_{\rm c}=\pm 1/\sqrt{3}$) and
$P_{\rm c}={\rm sgn}(u) (\sqrt{3}/2) F_{\rm c}$,
where subscript c denotes the critical point.
In addition, from the linear analysis around a critical point,
the velocity gradient near to the singularity is found to be
$du/d\tau |_{\rm c} = {\rm sgn} (u) (3/4)$.

We show an example of ``critical solutions'' in figure 3,
where the parameters are 
$\tau_{\rm c}=1$, $F_{\rm c}/(\pi I_{\rm s})=1$, 
and $J/(\pi I_{\rm s}/c^2)=1$.
As can be seen in figure 3,
one of the critical solutions (a solid curve)
has a positive velocity, and is an outward breeze solution,
which decelerates toward low optical depth.
Another (a dashed curve) has a negative velocity,
and is an inward settling solution,
which decelerates toward a high optical depth.
Both solutions decelerate in the direction of the flow
due to the radiation drag force.
However, neither satisfies the present boundary conditions
at the flow base and the flow top.
Indeed, in order for flow to be accelerated,
the velocity gradient $du/d\tau$ should always be negative,
but it is positive around a critical point for an outward solution.
Hence, the flow in the present framework
cannot pass through the critical point.
Thus, for the present purpose, 
such ``critical solutions'' are inadequate, and
the flow is always subsonic or supersonic
in the sense that the flow speed is always
less than or greater than $c/\sqrt{3}$.
And since we assume that the flow starts with $u=0$ at $\tau=\tau_0$,
we suppose a {\it subsonic} flow in the present paper:
$u_s < 1/\sqrt{2} \sim 0.707$.
We shall discuss this singularity problem later.

\begin{figure}
  \begin{center}
  \FigureFile(80mm,80mm){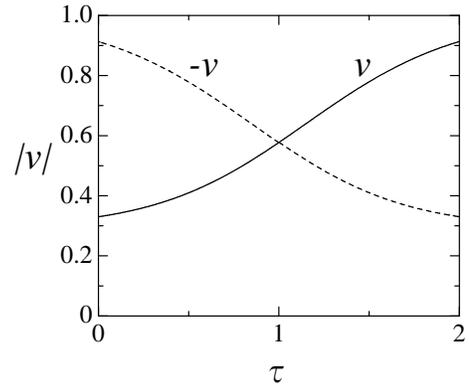}
  \end{center}
\caption{
Example of ``critical solutions''.
The parameters are 
$\tau_{\rm c}=1$, $F_{\rm c}/(\pi I_{\rm s})=1$, 
and $J/(\pi I_{\rm s}/c^2)=1$.
These critical solutions pass through the critical points
of the present problem,
but do not satisfy the present boundary conditions.
}
\end{figure}

\begin{figure}
  \begin{center}
  \FigureFile(80mm,80mm){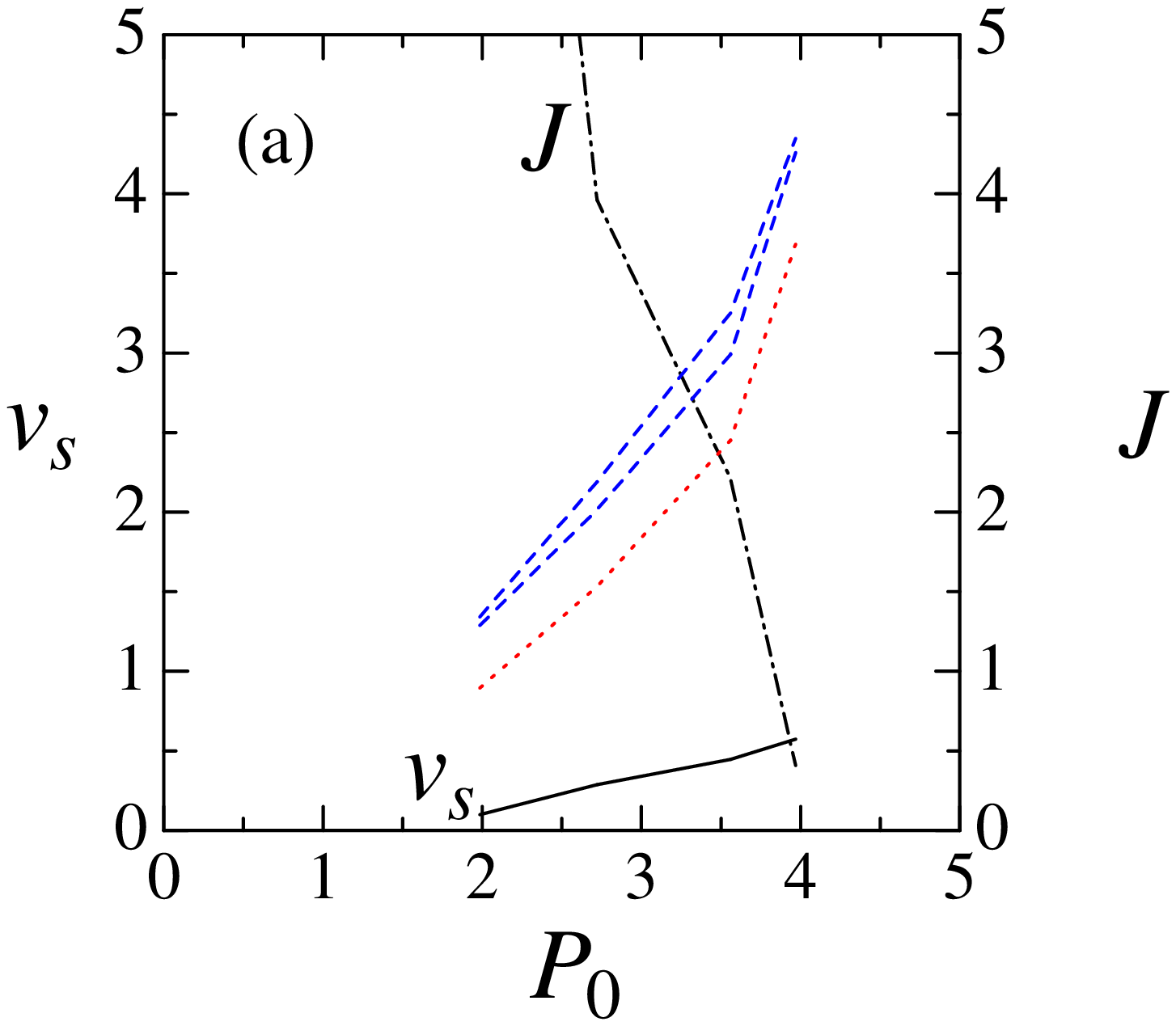}
  \end{center}
  \begin{center}
  \FigureFile(80mm,80mm){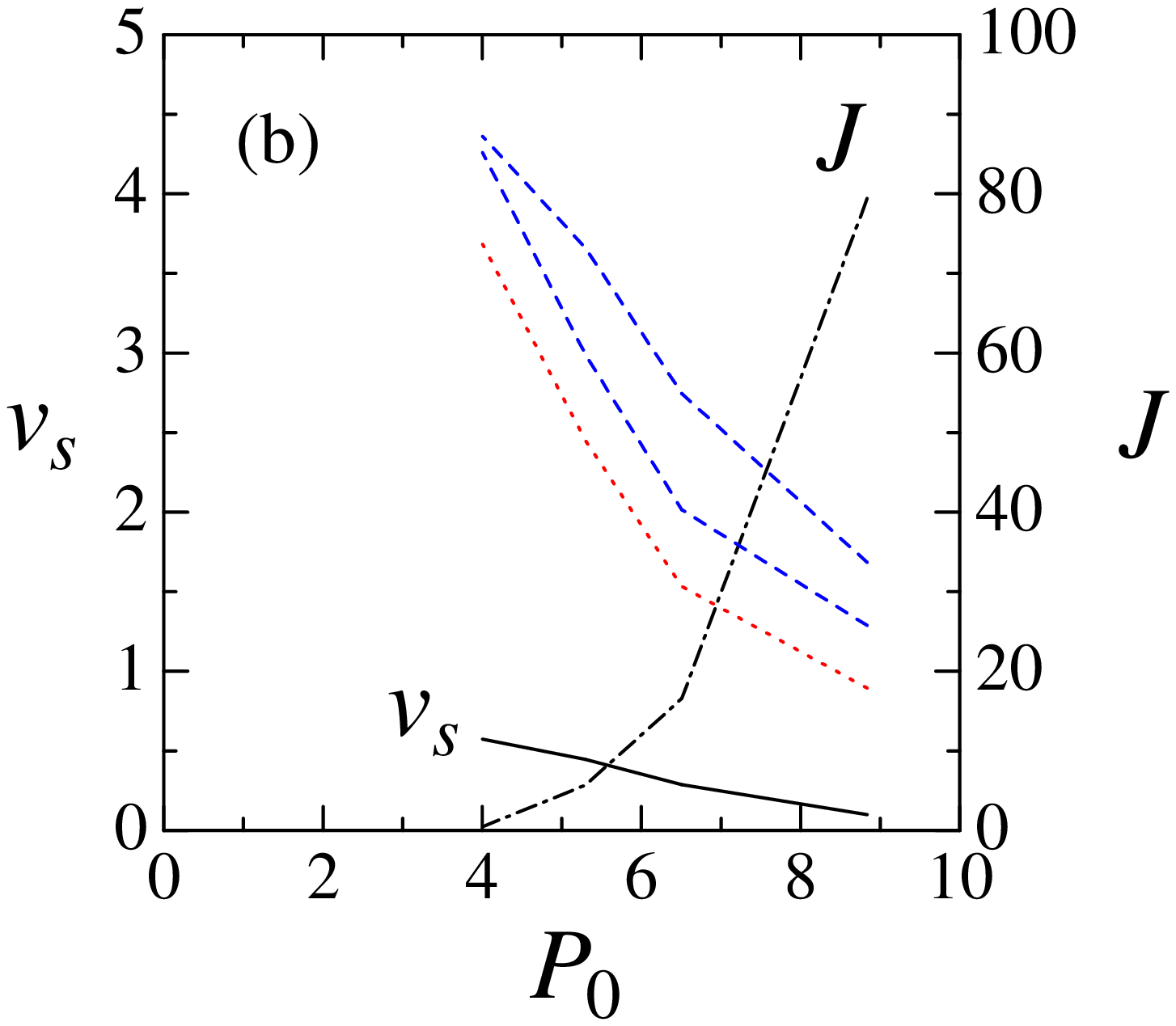}
  \end{center}
  \begin{center}
  \FigureFile(80mm,80mm){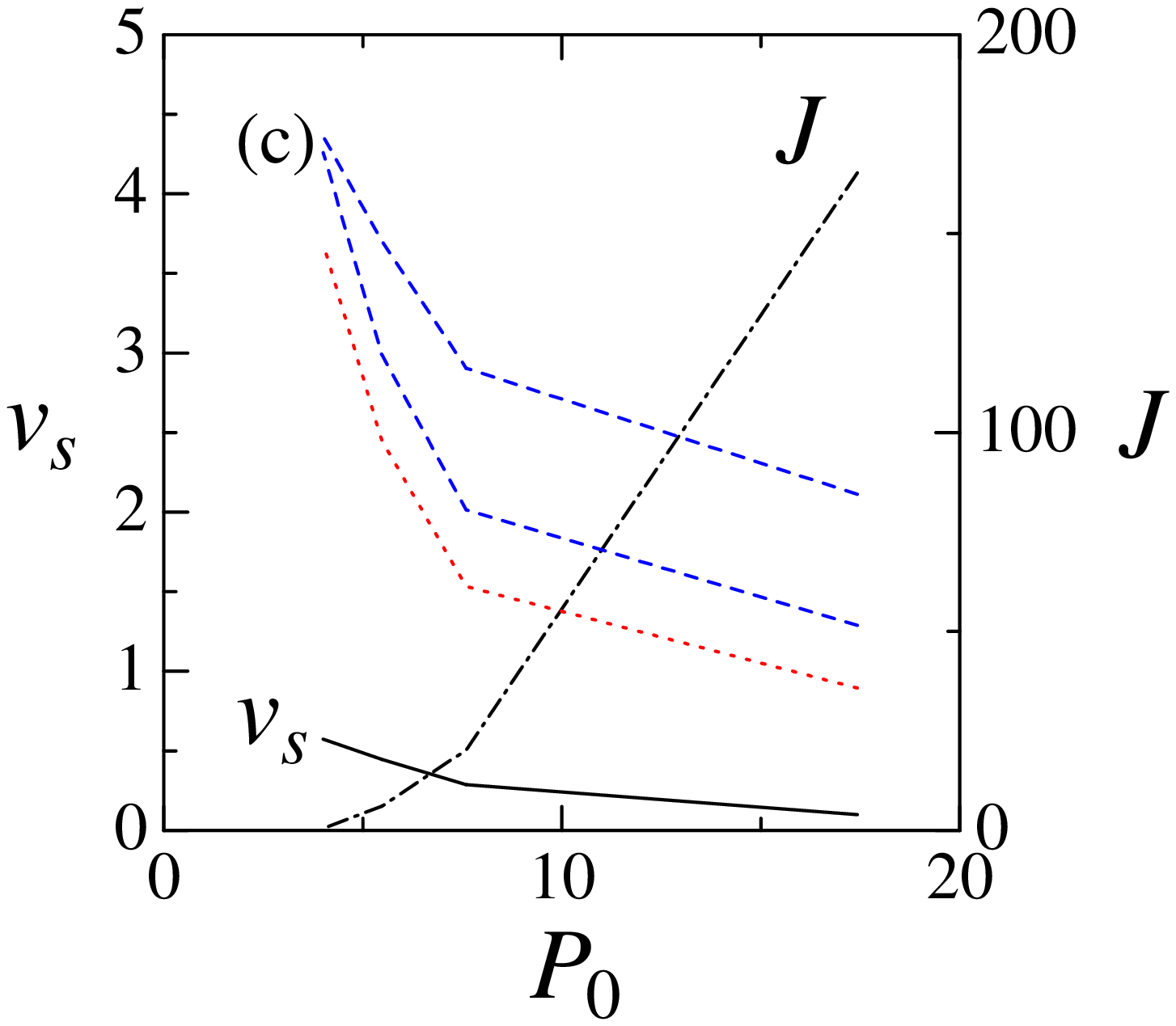}
  \end{center}
\caption{
Final velocity $v_{\rm s}$ at the flow top (solid curves),
the radiative flux $F_0$ at the flow base (upper dashed ones),
the radiative flux $F_{\rm s}$ at the flow top (lower dashed ones),
the radiation pressure $P_{\rm s}$ at the flow top (dotted ones),
and the mass-loss rate $J$ (chain-dotted ones),
as a function of $P_0$
for several values of $\tau_0$ at the flow base:
(a) $\tau_0=1$, (b) $\tau_0=10$, and (c) $\tau_0=100$.
The quantities are normalized in units of $c$ and $\pi I_{\rm s}$.
That is, the unit of $F$ and $cP$ is $\pi I_{\rm s}$
and the unit of $J$ is $\pi I_{\rm s}/c^2$.
The left-side scale of the ordinates is
for quantities except for $J$, while
the right-side is for $J$.
}
\end{figure}

\begin{figure}
  \begin{center}
  \FigureFile(80mm,80mm){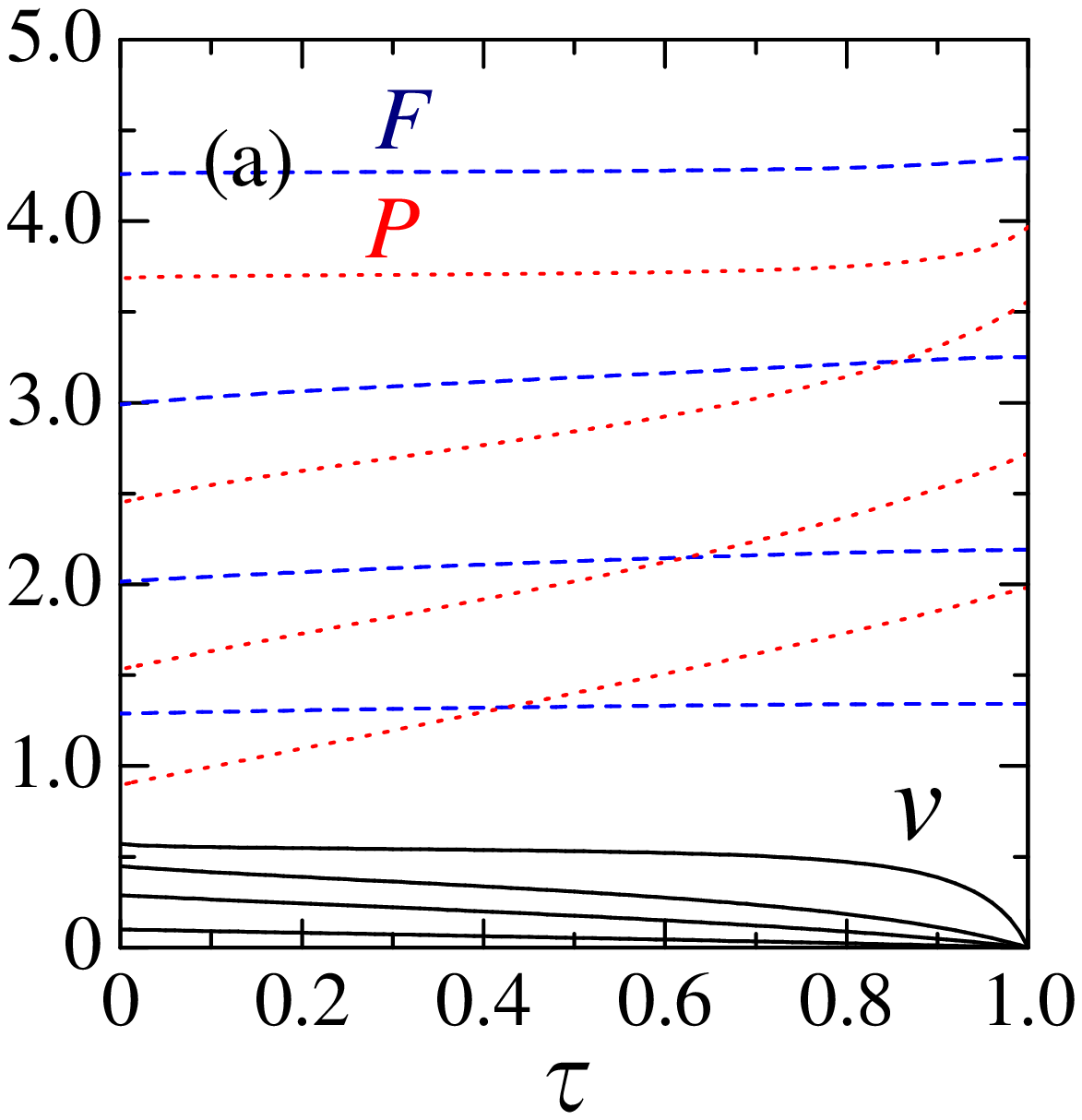}
  \end{center}
  \begin{center}
  \FigureFile(80mm,80mm){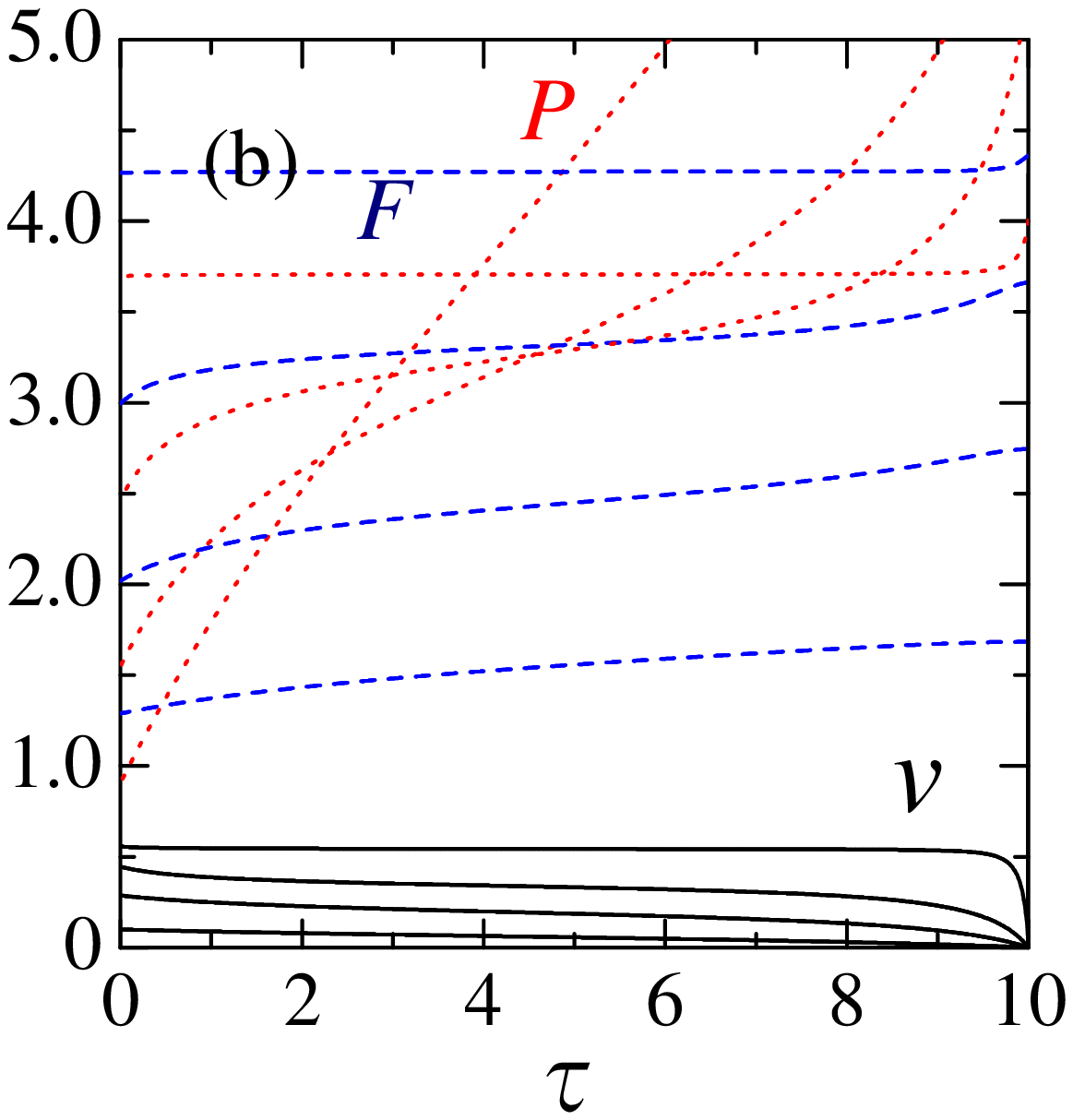}
  \end{center}
  \begin{center}
  \FigureFile(80mm,80mm){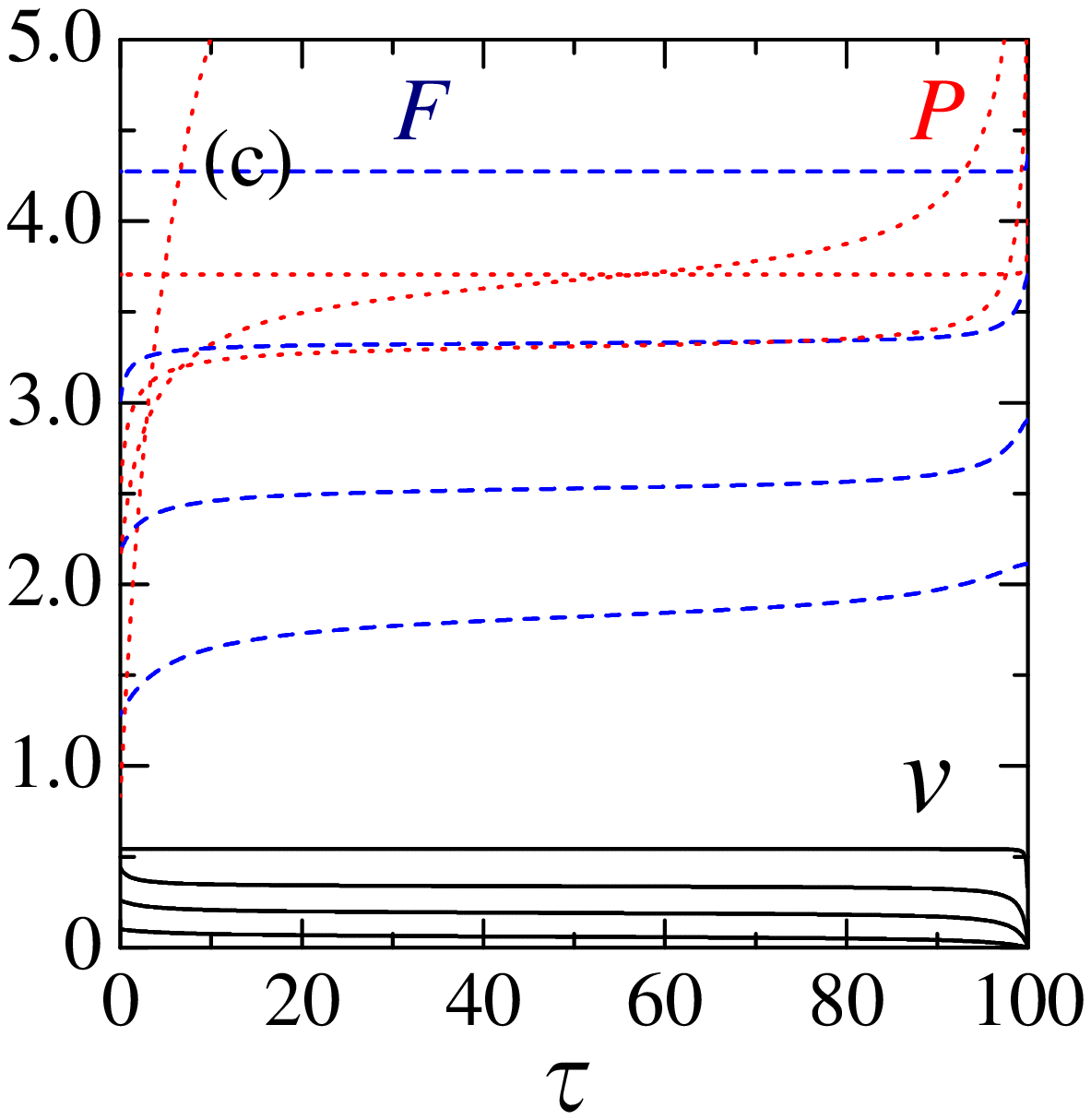}
  \end{center}
\caption{
Flow three velocity $v$ (solid curves), 
radiative flux $F$ (dashed curves),
and radiation pressure $P$ (dotted curves),
as a function of the optical depth $\tau$
for several values of $u_{\rm s}$ at the flow top
in a few cases of $\tau_0$.
The parameters are
(a) $\tau_0=1$, (b) $\tau_0=10$, and (c) $\tau_0=100$, and
 from bottom to top of $v$, $F$, and $P$,
the values of $u_{\rm s}$ are
0.1, 0.3, 0.5, and 0.7.
The quantities are normalized in units of $c$ and $\pi I_{\rm s}$.
That is, the unit of $F$ and $cP$ is $\pi I_{\rm s}$.
}
\end{figure}

\subsection{Subsonic Solutions}

Examples of the results for relativistic radiative flows
 in a luminous disk under the present boundary conditions
are shown in figures 4 and 5.

In figure 4
we show the final velocity $v_{\rm s}$ at the flow top (solid curves),
the radiative flux $F_0$ at the flow base (upper dashed ones),
the radiative flux $F_{\rm s}$ at the flow top (lower dashed ones),
the radiation pressure $P_{\rm s}$ at the flow top (dotted ones),
and the mass-loss rate $J$ (chain-dotted ones),
as a function of $P_0$
for several values of $\tau_0$ at the flow base.
The quantities are normalized in units of $c$ and $\pi I_{\rm s}$.
For example, the unit of $J$ is $\pi I_{\rm s}/c^2$.

As can be seen in figure 4,
as the radiative flux increases,
the final flow velocity at the flow top increases,
but the mass-loss rate decreases.
Moreover,
as can be seen in figure 4, 
and similar to the subrelativistic case (Fukue 2005b, c),
in order for flow to exist,
the radiation pressure $P_0$ at the flow base is restricted
to be within some range.
In the subrelativistic case without gravity and heating (Fukue 2005b),
the initial pressure $P_0$ is proved to be restricted within the range of
$2/3 < {cP_0}/{F_{\rm s}} < 2/3 +  \tau_0$.
In the present case,
the initial pressure is also restricted to be within a similar range,
but somewhat modified due to the relativistic effect:
i.e., the radiative flux $F_{\rm s}$ is no longer constant,
but depends on the flow final speed.
At the one limit of $P_0$,
the pressure gradient between the flow base and the top is maximum,
the loaded mass diverges, and the flow final speed becomes zero.
In the other limit of $P_0$,
the pressure gradient vanishes, 
the loaded mass becomes zero, and the flow final speed becomes high.

As already stated, in the present model,
the flow final speed is supposed to be within the range of
$0< u_{\rm s} < 1/\sqrt{2}$ (i.e., $0 < v_{\rm s}/c < 1/\sqrt{3}$).
As a result, using equations (\ref{Fs2}) and (\ref{Ps2}),
the radiative flux $F_{\rm s}$ and 
the radiation pressure $P_{\rm s}$ are restricted to be 
within the following ranges:
$1 < F_{\rm s}/(\pi I_{\rm s}) < 4.31$ and
$2/3 < cP_{\rm s}/(\pi I_{\rm s}) < 3.73$.
Finally, 
although the flow final speed is less than the photon sound speed
($v_{\rm s}/c < 1/\sqrt{3} \sim 0.577$),
the flow final speed can exceed the magic speed
above the luminous infinite disk
[$v_{\rm magic}/c = (4-\sqrt{7})/3 \sim 0.451$].

In figure 5
we show the flow three velocity $v$ (solid curves),
the radiative flux $F$ (dashed curves),
and the radiation pressure $P$ (dotted curves)
as a function of the optical depth $\tau$
for several values of $u_{\rm s}$ at the flow top
in a few cases of $\tau_0$.
The quantities are normalized in units of $c$ and $\pi I_{\rm s}$.

When the initial radiative flux $F_0$ at the flow base is large,
the flow is effectively accelerated, and
the flow final speed becomes large.
On the other hand,
when the pressure gradient between the flow base and the flow top is
large, the loaded mass $J$ becomes large (cf. figure 4).
The latter properties are somewhat complicated,
and should be explained much more.

When the optical depth $\tau_0$ is not so large (figures 4a and 5a),
the pressure gradient between the flow base and top is not so large.
In this case,
the pressure gradient becomes large,
as the pressure $P_0$ at the flow base becomes small.
This is the reason that the loaded mass becomes large,
since $P_0$ is small (figure 4a).
When the optical depth is large (figures 4b, 4c and 5b, 5c),
on the other hand,
the pressure gradient generally becomes large.
Hence,
the pressure gradient is large,
as the pressure $P_0$ at the flow base becomes large.
This is the reason that the loaded mass becomes large,
since $P_0$ is large (figures 4b and 4c).

Alternatively,
in the less-luminous, sub/non-relativistic limit with small $u_{\rm s}$,
the radiation fields are roughly expressed as
\begin{eqnarray}
   F &\sim& F_{\rm s},
\\
   cP &\sim& F_{\rm s}(2/3+\tau),
\end{eqnarray}
which are the usual Milne approximations (cf. Fukue 2005b, c).
Hence, in the less-luminous, small velocity case,
the pressure gradient between the flow top and base
becomes large, as the optical depth is large.
In the highly-luminous, relativistic limit with large $u_{\rm s}$,
on the other hand,
the flow velocity quickly reaches the equilibrium velocity,
where the right-hand side of equation (\ref{u}) or (\ref{beta})
vanishes, and becomes constant as
\begin{equation}
   \beta \sim \frac{2}{3}\frac{cP}{F}
       - \sqrt{ \left( \frac{2}{3}\frac{cP}{F} \right)^2 - \frac{1}{3} },
\end{equation}
where the radiative flux $F$ and the radiation pressure $P$
are also approximately constant 
according to equations (\ref{K}) and (\ref{L}).
For example, in the limiting case of $\beta=1/\sqrt{3}$,
$cP/F=\sqrt{3}/2$.
As a result,
the pressure gradient is small even for large optical depth.

In any case, the mass-loss rate $J$ is not given arbitrarily,
but is determined as an eigenvalue,
similar to the subrelativistic case (Fukue 2005b).

\section{Concluding Remarks}

In this paper 
we have examined radiative flow in a luminous disk,
while taking into account radiative transfer,
in a fully relativistic manner
(cf. Fukue 2005b, c for a subrelativistic regime).
The vertical velocity $v$, the radiative flux $F$, 
and the radiation pressure $P$
are numerically solved as a function of the optical depth $\tau$
for the cases without gravity and heating.
At the flow base (disk ``inside'') where the flow speed is zero,
the initial optical depth, the initial radiative flux,
and the initial radiation pressure are $\tau_0$,
$F_0$, and $P_0$, respectively;
in the usual accretion disk
these quantities are determined in terms of
the central mass, the mass-accretion rate, and the viscous process
as a function of radius.
At the flow top (disk ``surface'') where the optical depth $\tau$ is zero,
the radiation fields ($E_{\rm s}$, $F_{\rm s}$, and $P_{\rm s}$)
should coincide with those
above a {\it moving} photosphere with the final speed $v_{\rm s}$ 
with uniform intensity.
In order to match this boundary condition,
the mass-loss rate $J$ is determined as an eigenvalue.

One of the relativistic manifestations is
this boundary condition at the flow top.
The radiation fields are not for the static flat source,
but should be for the moving flat source.
As a result,
it is found that the magic speed above a moving photosphere
can exceed that above a static photosphere ($\sim 0.45~c$).

Another relativistic manifestation is
the existence of the singular point in the equation,
which is related to the sound speed of the relativistic (photon) gas.
Since the ``critical solutions'' for the present problem
do not satisfy the present boundary conditions,
we obtained {\it subsonic} solutions,
which is always less than the relativistic sound speed of $c/\sqrt{3}$.
Nevertheless,
we can find the relativistic solution,
whose final speed is greater than
the magic speed above a static source.

The appearance of this singularity may come from
the closure relation for the quantities of radiation fields,
which assumes the usual Eddington approximation
in the comoving frame (Kato et al. 1998).
That is, in the comoving frame we assume
\begin{equation}
   P^{ij}_0 = \frac{\delta^{ij}}{3}E_0,
\label{P0E0}
\end{equation}
where subscript 0 denotes the quantities in the comoving frame.
The appearance of the singularity suggests that
the Eddington approximation would be violated
in the relativistic flow,
whose velocity is greater than $c/\sqrt{3}$.
This may be because
the diffusion speed in the comoving frame
cannot exceed $c/\sqrt{3}$, and/or
the diffusion would not be isotropic in the comoving frame.
Indeed, if we assume that
the diffusion in the comoving frame is no longer isotropic,
but the factor $1/3$ in equation (\ref{P0E0}) is variable,
similar to the usual variable Eddington factor,
we can formally extinguish the singularity.

The plane-parallel assumption may also play a role
in the existence of the singularity,
as well as the existence of the radiation drag force,
which is related to the closure approximation 
transformed to the inertial frame.
The nature of critical points should be examined more carefully.
In any case,
the existence of the singularity does not mean
that the flow velocity cannot exceed $c/\sqrt{3}$,
but merely suggests a possible failure of the formalism
adopted in the present analysis.

The relativistic radiative flow investigated in the present paper
must be quite {\it fundamental} to
accretion disk physics, astrophysical jet formation,
gamma-ray bursts, etc.,
although the present paper is only the first step and
there are many simplifications at the present stage.

At first, in a rigorous sense,
the present situations, such as a plane-parallel assumption,
are valid only in very specific circumstances.
Indeed, when the accretion rate exceeds the critical one,
the disk would puff up to be geometrically thick,
and the plane-parallel approximation would be violated.
In a rough sense, however,
the present situations could be approximately valid in such a thick disk,
as long as the gas motion and the radiative flux are almost vertical.
If the gas motion is not vertical, but expands like a spherical flow,
the present treatment must be reexamined.
In contrast to the plane-parallel case considered in the present paper,
the spherically expanding case is also of great interest to us,
and should be examined in the future.

In the present paper, we have assumed that 
the opacity is frequency-independent (gray), and therefore
the radiation pressure force is independent of the velocity.
In a cold gas, however, atomic absorption opacities become important.
Such opacities are frequency-dependent, and therefore
the radiation pressure force would be a function
of the Doppler shift between the main absorption troughs
and the peak of the black body spectrum from the disk.
Hence, the radiation energy flux $L$ would generally be
a function of the velocity $u$.
In addition, in such a case of atomic opacities,
the super-Eddington radiation-pressure force on the flow
may arise from a larger absorption opacity.
Since the acceleration nature would be drastically changed,
such a {\it line-driven radiative flow} must be carefully examined.

Furthermore, we ignored the gravitational field
produced by the central object.
This means that the flow considered in the present paper
would correspond to normal plasmas in the super-Eddington disk,
pair plasmas in the sub-Eddington disk, or
dusty plasmas in the luminous disk.
In other cases, or even for normal plasmas in the super-Eddington disk,
the gravitational field would affect the flow properties
(cf. Fukue 2005c for a subrelativistic case).
In particular, since the gravitational field in the vertical direction
is somewhat complicated (e.g., Fukue 2002, 2004),
the influence of the gravitational field is important.

We also ignored the gas pressure.
This means that the radiation field is sufficiently intense.
In general cases, where the gas pressure is considered,
there usually appear sonic points (e.g., Fukue 2002, 2004),
and the flow is accelerated from subsonic to supersonic.

In the high-energy regime, including relativistic flow and hot gas,
the Compton heating and cooling would be important,
although we have dropped such an effect.
If the Compton heating of the gas by intense radiation fields works,
the gas might be radiatively heated to very high temperatures,
and the atomic absorption opacity could be significantly reduced.
Such a radiative effect in the energy equation
should also be examined in future work.

Finally,
it is left as an open question
whether the singularity appearing in the present paper
can be passed through or whether
the relativistic radiation transfer formalism
should be improved.

\vspace*{1pc}

The authors would like to thank Professors S. Kato and S. Inagaki
for their useful comments and discussions.
Valuable comments from an anonymous referee are also gratefully acknowledged.
This work has been supported in part
by a Grant-in-Aid for Scientific Research (15540235 J.F.) 
of the Ministry of Education, Culture, Sports, Science and Technology.


\end{document}